\documentclass[natbib]{svjour3}
\usepackage{graphicx}

\begin{document}

\title{Coupling-Induced Population Synchronization in An Excitatory Population of Subthreshold Izhikevich Neurons}
\author{Sang-Yoon Kim  \and Woochang Lim}

\institute{S.-Y. Kim \at
              Research Division, LABASIS Corporation, Chunchon, Gangwon-Do 200-702, Korea
              \and
              W. Lim (corresponding author) \at
              Department of Science Education, Daegu National University of Education, Daegu 705-715, Korea
              Tel.: +82-53-620-1348 \\
              Fax: +82-53-620-1525 \\
              \email{woochanglim@dnue.ac.kr}
}


\maketitle

\begin{abstract}
We consider an excitatory population of subthreshold Izhikevich neurons which exhibit noise-induced firings. By varying the coupling strength $J$,
we investigate population synchronization between the noise-induced firings which may be used for efficient cognitive processing such as sensory perception, multisensory binding, selective attention, and memory formation. As $J$ is increased, rich types of population synchronization (e.g.,  spike, burst, and fast spike synchronization) are found to occur. Transitions between population synchronization and incoherence are well described
in terms of an order parameter $\cal{O}$. As a final step, the coupling induces oscillator death (quenching of
noise-induced spikings) because each neuron is attracted to a noisy equilibrium state. The oscillator death leads to a transition from firing to non-firing states at the population level, which may be well described in terms of the time-averaged population spike rate $\overline{R}$. In addition to the statistical-mechanical analysis using $\cal{O}$ and $\overline{R}$, each population and individual state are also characterized by using the techniques of nonlinear dynamics such as the raster plot of neural spikes, the time series of the membrane potential, and the phase portrait. We note that population synchronization of noise-induced firings may lead to emergence of synchronous brain rhythms in a noisy environment, associated with diverse cognitive functions.
\end{abstract}


\section{Introduction}
\label{sec:INT}

Recently, much attention has been paid to brain rhythms observed in scalp electroencephalogram (EEG) and local field potentials (LFP) with electrodes inserted into the brain \citep{Buz}. These brain rhythms emerge via synchronization between individual firings in neural circuits. Population synchronization between neural firing activities may be used for efficient sensory and cognitive processing such as sensory perception, multisensory integration, selective attention, and working memory \citep{W_Review}. Many recent works have been investigated in diverse views of population synchrony \citep{PopSync3,PopSync4,PopSync2,PopSync1,PopSync5,PopSync6, PopSync7,PopSync8,PopSync9,PopSync10,PopSync11,PopSync12,PopSync13}. This kind of population synchronization is also correlated with pathological rhythms associated with neural diseases \citep{TW}. Here, we are interested in these synchronous brain rhythms. Population synchronization has been intensively investigated in neural circuits composed of spontaneously firing suprathreshold neurons exhibiting clock-like regular discharges \citep{W_Review, Wang}. For this case, population synchronization may occur via cooperation of regular firings of suprathreshold self-firing neurons. In contrast to the suprathreshold case, the case of subthreshold neurons has received little attention. For an isolated single case, a subthreshold neuron cannot fire spontaneously; it can fire only with the help of noise. Here we are interested in population synchronization between complex noise-induced firings of subthreshold neurons which exhibit discharges like Geiger counters. Recently, noise-induced population synchronization was studied by varying the noise intensity observed in a population of subthreshold neurons, and thus collective coherence between noise-induced firings has been found to occur in an intermediate range of noise intensity \citep{CR,Kim1A,Kim1B,Lim1,Lim2,Kim2,Kim3}.

In this paper, we investigate coupling-induced population synchronization which leads to emergence of synchronous brain rhythms by varying the coupling strength $J$ in an excitatory population of globally coupled subthreshold Izhikevich neurons, and thus rich types of population synchronization are found to emerge. As an element in our coupled neural system, we choose a simple Izhikevich neuron \citep{Burst1,Izhi1,Izhi2,Izhi3} which is as biologically plausible as the Hodgkin-Huxley model \citep{HH}, yet as computationally efficient as the integrate-and-fire model \citep{IF1,IF2}. These Izhikevich neurons interact via excitatory AMPA synapses in our computational study. For small $J$ individual neurons fire spikings independently, and thus the population state is incoherent. However, when passing a lower threshold $J_l^*$, population spike synchronization occurs because the coupling stimulates coherence between noise-induced spikings. As in globally-coupled chaotic systems, this kind of transition between population synchronization and incoherence may be well described in terms of an order parameter $\cal{O}$ \citep{Order1A,Order1B,Order1C,Order2}; in our case, the time-averaged fluctuation of the population-averaged membrane potential plays the role of $\cal{O}$. As $J$ is further increased and passes another threshold $J_b^*$, noise-induced burstings appear in individual membrane potentials, and population burst synchronization also emerges. In contrast to spiking activity, bursting activity alternates between a silent phase and an active phase of repetitive spikings \citep{Burst1,Burst2,Burst3}. This type of burstings are known to play the important roles in neural communication \citep{Izhi1,Izhi2,Izhi3}. As $J$ continues to increase, the length of active phase in individual bursting potential increases, and eventually a transition from bursting to fast spiking occurs at a threshold $J_{fs}^*$. Consequently, breakup of population burst synchronization occurs and incoherent states appear because individual fast spikings keep no pace with each other. However, as $J$ is further increased, coupling stimulates population synchronization between fast spikings in a range of $J_{h1}^* < J < J_{h2}^*$. For $J > J_{h2}^*$ population states become incoherent and slow spikings appear in individual membrane potentials. As a final step, when $J$ passes a high threshold $J_{od}^*$, coupling induces oscillator death (i.e., quenching of noise-induced slow spikings of individual neurons) because each neuron is attracted to a noisy equilibrium state. This stochastic oscillator death in the presence of noise \citep{SOD1,SOD2} is in contrast to the deterministic oscillator death occurring in the absence of noise \citep{OD}. At the population level, a transition from firing to non-firing states results from stochastic oscillator death. We also characterize the firing-nonfiring transition in terms of the time-averaged population spike rate $\overline{R}$ which plays a role similar to that of the order parameter $\cal{O}$ for the incoherence-coherence transition. In addition to the statistical-mechanical analysis using $\cal{O}$ and $\overline{R}$, these diverse population and individual states are well characterized by using the techniques of nonlinear dynamics such as the raster plot of spikes, the time series of the membrane potential, and the phase portrait.

\section{Subthreshold Izhikevich Neuron Model}
\label{sec:MODEL}

We consider an excitatory population of $N$ globally-coupled subthreshold neurons. As an element in our coupled neural system, we choose the simple Izhikevich neuron model which is not only biologically plausible, but also computationally efficient \citep{Burst1,Izhi1,Izhi2,Izhi3}. The population dynamics in this neural network is governed by the following set of ordinary differential equations:
\begin{eqnarray}
 \frac{dv_i}{dt} &=& f(v_i)- u_i +I_{DC} +D \xi_{i} -I_{syn,i}, \label{eq:CIZA} \\
\frac{du_i}{dt} &=& a (bv_i - u_i), \label{eq:CIZB} \\
\frac{ds_i}{dt}&=& \alpha s_{\infty}(v_i) (1-s_i) - \beta s_i, \;\;\; i=1, \cdots, N, \label{eq:CIZC}
\end{eqnarray}
with the auxiliary after-spike resetting:
\begin{equation}
{\rm if~} v_i \geq v_p,~ {\rm then~} v_i \leftarrow c~ {\rm and~} u_i \leftarrow u_i + d, \label{eq:RS}
\end{equation}
where
\begin{eqnarray}
f(v_i) &=& 0.04 v_i^2 + 5 v_i + 140, \label{eq:CIZD} \\
I_{syn,i} &=& \frac{J}{N-1} \sum_{j(\ne i)}^N s_j(t) (v_i -
V_{syn}), \label{eq:CIZE} \\
s_{\infty} (v_i) &=& 1/[1+e^{-(v_i-v^*)/\delta}]. \label{eq:CIZF}
\end{eqnarray}

We note that $f(v)$ of Eq.~(\ref{eq:CIZD}) was obtained by fitting the spike initiation dynamics of cortical neurons so that the membrane potential $v$ has mV scale and the time $t$ has ms scale \citep{Izhi1,Izhi2,Izhi3}. The state of the $i$th neuron at a time $t$ is characterized by three dimensionless state variables: the membrane potential $v_i$, the recovery variable $u_i$ representing the activation of the $K^+$ ionic current and the inactivation of the $Na^+$ ionic current, and the synaptic gate variable $s_i$ denoting the fraction of open synaptic ion channels. After the spike reaches its apex $v_p$ (=30 mV), the membrane voltage and the recovery variable are reset according to Eq.~(\ref{eq:RS}). There are four dimensionless parameters, $a, b, c,$ and $d$ representing the time scale of the recovery variable $u$, the sensitivity of $u$ to the subthreshold fluctuations of $v$, and the after-spike reset value of $v$ and $u$, respectively. Tuning the four parameters, the Izhikevich neuron model may produce 20 of the most prominent neuro-computational features of cortical neurons \citep{Izhi1,Izhi2,Izhi3}. Unlike Hodgkin-Huxley-type conductance-based models, the Izhikevich model matches neuronal dynamics instead of matching neuronal electrophysiology.

Each Izhikevich neuron is stimulated by the common DC current $I_{DC}$ and an independent Gaussian white noise $\xi$ [see the 2nd and 3rd terms in Eq.~(\ref{eq:CIZA})] satisfying $\langle \xi_i(t) \rangle =0$ and $\langle \xi_i(t)~\xi_j(t') \rangle = \delta_{ij}~\delta(t-t')$, where $\langle\cdots\rangle$ denotes the ensemble average. The noise $\xi$ is a parametric one which randomly perturbs the strength of the applied current $I_{DC}$, and its intensity is controlled by the parameter $D$. The last term in Eq.~(\ref{eq:CIZA}) represents the coupling of the network. Each neuron is connected to all the other ones through global couplings via excitatory AMPA synapses. $I_{syn,i}$ of Eq.~(\ref{eq:CIZE}) represents such synaptic current injected into the $i$th neuron. Here the coupling strength is controlled by the parameter $J$ and $V_{syn}$ is the synaptic reversal potential. We use $V_{syn}=10$ mV for the excitatory synapse. The synaptic gate variable $s$ obeys the 1st order kinetics of Eq.~(\ref{eq:CIZC}) \citep{GR1,GR2}. Here, the normalized concentration of synaptic transmitters, activating the synapse, is assumed to be an instantaneous sigmoidal function of the membrane potential with a threshold $v^*$ in Eq.~(\ref{eq:CIZF}), where we set $v^*=0$ mV and $\delta=2$ mV. The transmitter release occurs only when the neuron emits a spike (i.e., its potential $v$ is larger than $v^*$). For the excitatory glutamate synapse (involving the AMPA receptors), the synaptic channel opening rate, corresponding to the inverse of the synaptic rise time $\tau_r$, is $\alpha=10$ ${\rm ms}^{-1}$, and the synaptic closing rate $\beta$, which is the inverse of the synaptic decay time $\tau_d$, is $\beta=0.5$ ${\rm ms}^{-1}$ \citep{BK1,BK2}.

\begin{figure}
\includegraphics[width=0.8\columnwidth]{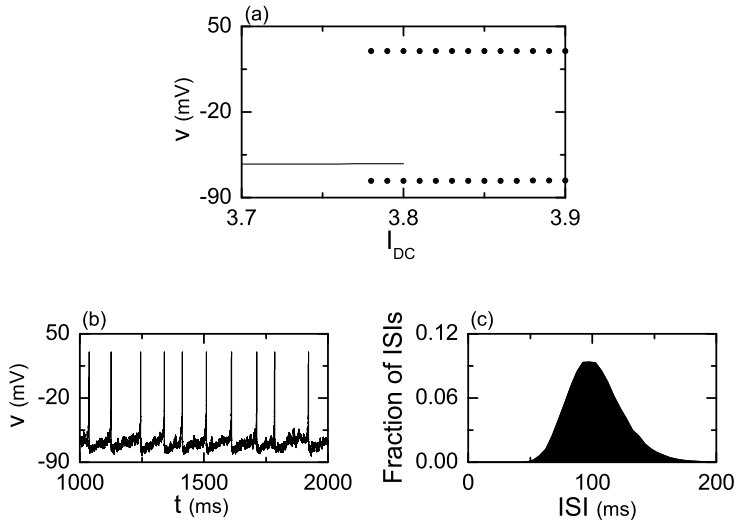}
\caption{(a) Bifurcation diagram in the single type-II regular-spiking Izhikevich neuron for $D=0$. Solid line denotes a stable equilibrium point. Maximum and minimum values of $v$ for the spiking state are represented by solid circles. (b) Time series of the membrane potential $v(t)$ and (c) the interspike interval histogram in the single subthreshold Izhikevich neuron for $I_{DC}=3.6$ and $D=3.0$.
}
\label{fig:Single}
\end{figure}

Here we consider the case of regular-spiking cortical excitatory neurons for $a=0.02$, $b=0.2$, $c=-65$, and $d=8$. Depending on the system parameters, the Izhikevich neurons may exhibit either type-I or type-II excitability \citep{Burst3,Izhi1,Izhi2,Izhi3}; for the case of type-I (type-II) neurons, the firing frequency begins to increase from zero (non-zero finite value) when $I_{DC}$ passes a threshold \citep{Ex1,Ex2}. For our case, a deterministic Izhikevich neuron (for $D=0$) exhibits a jump from a resting state (denoted by solid line) to a spiking state (denoted by solid circles) via a subcritical Hopf bifurcation for $I_{DC,h}^* =3.80$ by absorbing an unstable limit cycle born via a fold limit cycle bifurcation for $I_{DC,l}^* = 3.78$, as shown in Fig.~\ref{fig:Single}(a). Hence, the Izhikevich neuron shows the type-II excitability because it begins to fire with a non-zero frequency that is relatively insensitive to the change in $I_{DC}$. Throughout this paper, we consider a subthreshold case of $I_{DC}=3.6$. An isolated subthreshold Izhikevich neuron cannot fire spontaneously without noise. Figures \ref{fig:Single}(b) and \ref{fig:Single}(c) show a time series of the membrane potential $v$ of a subthreshold neuron and the interspike interval histogram for $D=3.0$. Complex noise-induced subthreshold oscillations and spikings with irregular interspike intervals appear. Population synchronization is investigated in an excitatory population of these subthreshold Izhikevich neurons coupled via AMPA synapses. Hereafter, we fix the value of the noise intensity as $D=3.0$. Numerical integration of the governing equations  (\ref{eq:CIZA})-(\ref{eq:CIZC}) is done using the Heun method \citep{SDE} (with the time step $\Delta t=0.01$ ms). For each realization of the stochastic process in Eqs.~(\ref{eq:CIZA})-(\ref{eq:CIZC}), we choose a random initial point $[v_i(0),u_i(0),s_i(0)]$ for the $i$th $(i=1,\dots, N)$ neuron with uniform probability in the range of $v_i(0) \in (-70,30)$, $u_i(0) \in (-10,-6)$, and $s_i(0) \in (0,1)$.

\section{Coupling-Induced Population Synchronization}
\label{sec:PS}

\begin{figure}
\includegraphics[width=0.8\columnwidth]{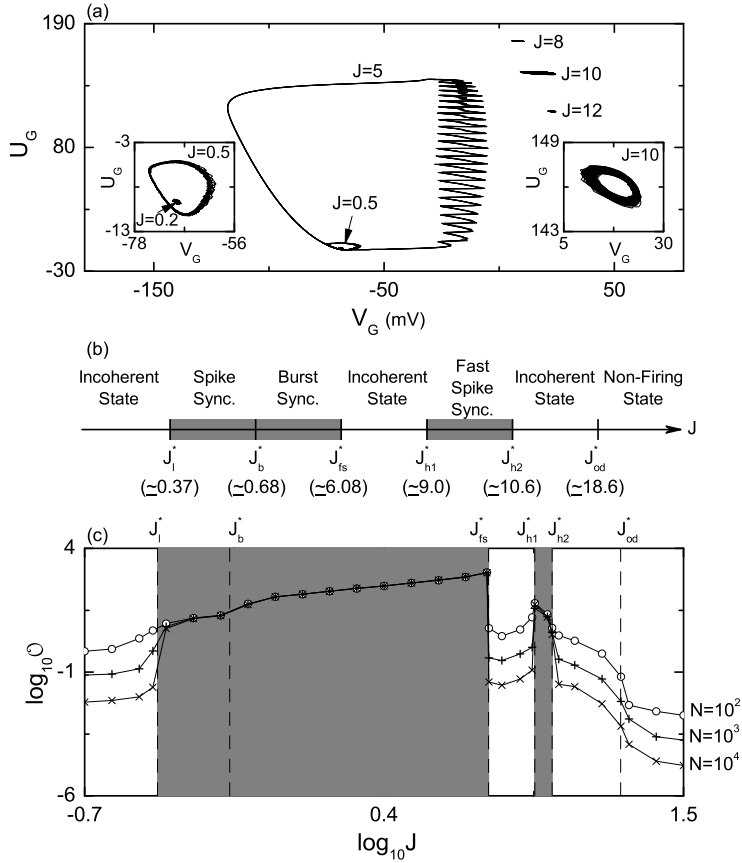}
\caption{Population coherent and incoherent states in $N$ globally-coupled excitatory subthreshold Izhikevich neurons for $I_{DC}=3.6$ and $D=3.0$. (a) Phase portraits of the population states in the $V_G-U_G$ plane for $N=10^3$. (b) Schematic diagram of populations states on the $J$ axis (c) Transition between coherence and incoherence: plots of $\log_{10}{\cal{O}}$ versus $\log_{10}J$ for $N=10^2$, $10^3$, and $10^4$. Spike, burst, and fast spike synchronizations occur in the gray regions in (b) and (c).
}
\label{fig:Order}
\end{figure}

By varying the coupling strength $J$, we investigate population synchronization, via which synchronous brain rhythms emerge, by using diverse techniques of statistical mechanics and nonlinear dynamics. Emergence of population synchronization may be described by the population-averaged membrane potential $V_G$ (corresponding to the global potential) and the global recovery variable $U_G$,
\begin{equation}
 V_G (t) = \frac {1} {N} \sum_{i=1}^{N} v_i(t)~~~{\rm and}~~~
 U_G (t) = \frac {1} {N} \sum_{i=1}^{N} u_i(t).
\label{eq:GPOT}
\end{equation}
Figure \ref{fig:Order}(a) shows rich phase portraits of the representative coherent and incoherent population states in the $V_G-U_G$ phase plane. Population synchronization appears on noisy limit cycles for $J=0.5, 5$ and $10$, while incoherent states occur on noisy equilibrium points for $J=0.2,8$, and $12$. Particularly, for $J=5$ population burst synchronization emerges on a noisy hedgehoglike limit cycle; spines and body correspond to active and silent phases of the bursting activity, respectively. A schematic phase diagram of these population states on the $J$ axis is shown in Fig.~\ref{fig:Order}(b). Transitions between incoherent and coherent states may be well described in terms of the order parameter. For our case, the mean square deviation of the global potential $V_G(t)$ (i.e., time-averaged fluctuations of $V_G(t)$),
\begin{equation}
{\cal{O}} \equiv \overline{(V_G(t) - \overline{V_G(t)})^2},
 \label{eq:Order}
\end{equation}
plays the role of an order parameter, where the overbar represents the time averaging. Here, we discard the first time steps of a stochastic trajectory as transients during $10^3$ ms, and then we numerically compute $\cal{O}$ by following the stochastic trajectory for $3 \times 10^4$ ms when $N=10^2, 10^3,$ and $10^4$. For the coherent (incoherent) state, the order parameter $\cal{O}$ approaches a nonzero (zero) limit value in the thermodynamic limit of $N \rightarrow \infty$. Figure \ref{fig:Order}(c) shows a plot of the order parameter versus the coupling strength. For $J < J^*_l$ $(\simeq 0.37)$, the order parameter $\cal{O}$ tends to zero as $N \rightarrow \infty$, and hence incoherent states exist. As $J$ passes the lower threshold $J^*_l$, a coherent transition to spike synchronization occurs because the coupling stimulates coherence between noise-induced spikings. Thus, spike synchronization appears for $J_l^* < J < J^*_b$ $(\simeq 0.68)$. However, when passing another threshold $J^*_b$, individual neurons exhibit noise-induced burstings and population burst synchronization occurs. As $J$ is further increased and passes a threshold $J_{fs}^*$ $(\simeq 6.08$), a transition from bursting to fast spiking occurs in individual potentials and the burst synchronization breaks up because individual fast spikes keep no pace with each other. Thus, for $J > J_{fs}^*$ the order parameter $\cal{O}$ tends to zero as $N \rightarrow \infty$, and incoherent states appear. However, with further increase in $J$, coupling-induced fast spike synchronization occurs in a range of $J^*_{h1} (\simeq 9.0) < J < J_{h2}^* (\simeq 10.6)$. For $J > J^*_{h2}$ incoherent states reappear as shown in Fig.~\ref{fig:Order}(c), and individual neurons exhibit slow spikings. As a final step, when passing a high threshold $J_{od}^*$ $(\simeq 18.6)$, coupling induces stochastic oscillator death (i.e., cessation of noise-induced slow spikings) because each neuron is attracted to a noisy equilibrium state. This stochastic oscillator death leads to a transition from  firing to non-firing state at the population level. In this way, three kinds of population synchronization (i.e., spike, burst, and fast spike synchronization) emerge in the gray regions of Figs.~\ref{fig:Order}(b) and \ref{fig:Order}(c).

\begin{figure}
\includegraphics[width=0.8\columnwidth]{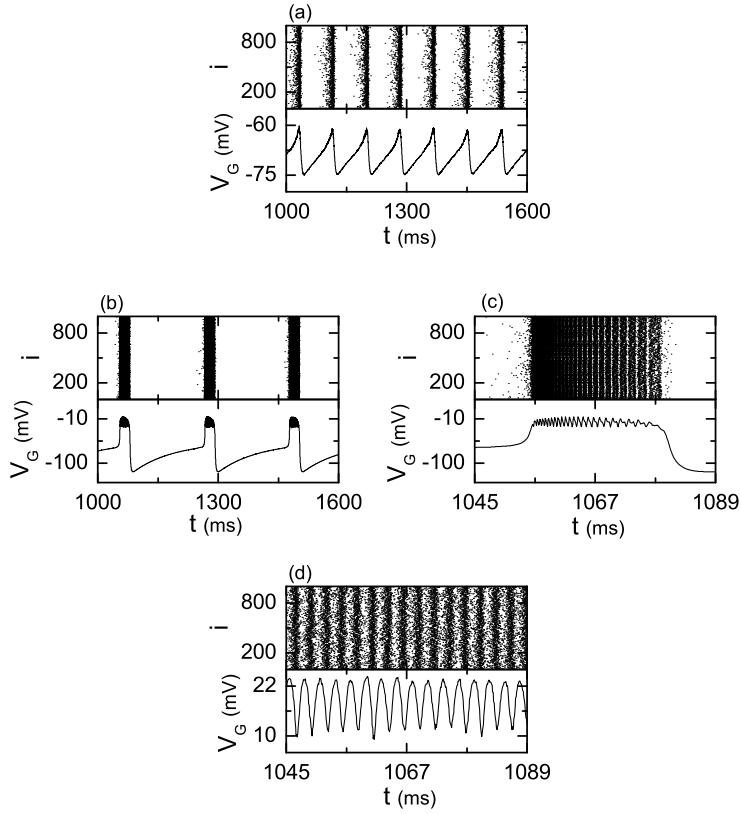}
\caption{Three types of population synchronization in $N$ $(=10^3)$ globally-coupled excitatory subthreshold Izhikevich neurons
for $I_{DC}=3.6$ and $D=3.0$. Raster plots of spikes and time series of the global potential $V_G(t)$ for (a) spike synchronization ($J=0.5$), (b1) and (b2) burst synchronization ($J=5$), and (c) fast spike synchronization ($J=10$).
}
\label{fig:Population}
\end{figure}

We present population synchronization clearly in terms of the raster plots of spikes and the time series of the global potential $V_G$. The first spike synchronization appears in a range of $J_{l}^* < J < J_b^*$. An example for $J=0.5$ is shown in Fig.~\ref{fig:Population}(a). Stripes (composed of spikes), indicating population synchronization, appear regularly with the mean time interval $(\simeq 83.6$ ms) in the raster plot, and $V_G$ shows a small-amplitude negative-potential population rhythm with frequency $f_g =12$ Hz. The second burst synchronization occurs in a range of $J_{b}^* < J < J_{fs}^*$. Figure \ref{fig:Population}(b1) shows bursting synchronization for $J=5$. Clear burst bands, composed of stripes of spikes, appear successively at nearly regular time intervals $(\simeq 210.7 $ ms) in the raster plot, and the corresponding global potential $V_G$ exhibits a large-amplitude bursting rhythm with $f_g=4.75$ Hz. In contrast to spiking rhythm in Fig.~\ref{fig:Population}(a), much more hyperpolarization occurs in the bursting rhythm. For a clear view, magnifications of a single burst band and $V_G$ are given in Fig.~\ref{fig:Population}(b2). For this kind of burstings, burst synchronization refers to a temporal relationship between the active phase onset or offset times of bursting neurons, while spike synchronization characterizes a temporal relationship between spikes fired by different bursting neurons in their respective active phases \citep{Sync}. In addition to burst synchronization, spike synchronization also occurs in each burst band, as shown in Fig.~\ref{fig:Population}(b2); as we go from the onset to the offset times, wider stripes appear in the burst band. Hence, this kind of burst synchronization occurs on a hedgehoglike limit cycle [see Fig.~\ref{fig:Order}(a)], and $V_G$ exhibits bursting activity like individual potentials. Finally, the third fast spike synchronization emerges in a range of $J_{h1}^* < J < J_{h2}^*$. An example for $J=10$ is shown in Fig.~\ref{fig:Population}(c). In contrast to Fig.~\ref{fig:Population}(a), stripes appear successively at short time intervals $(\simeq 2.8$ ms) in the raster plot, and $V_G$ shows a small-amplitude positive-potential fast rhythm with $f_g \simeq 356$ Hz.

\begin{figure}
\includegraphics[width=0.8\columnwidth]{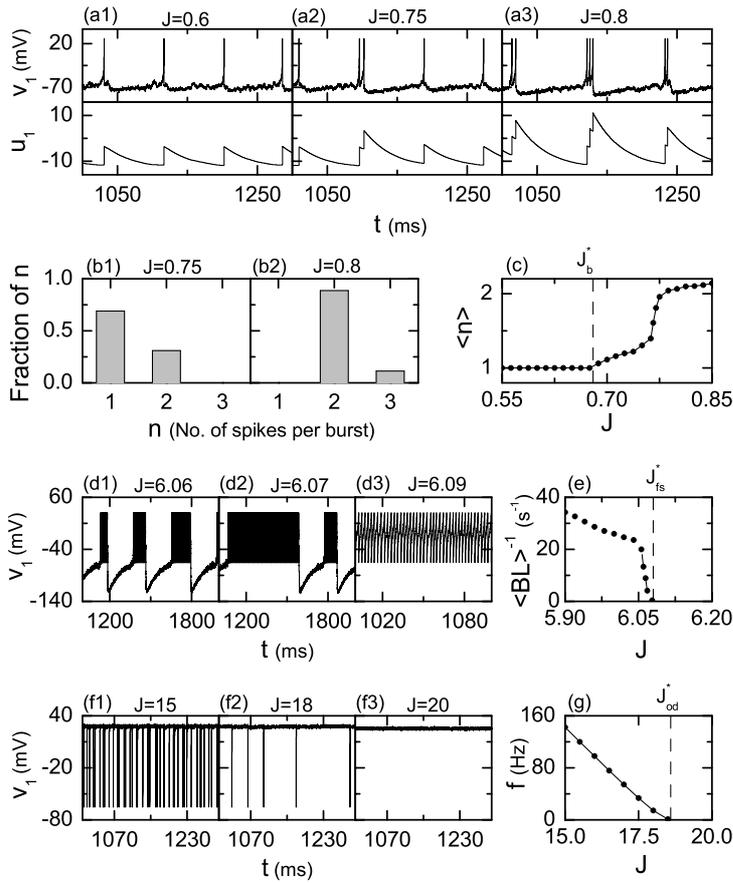}
\caption{Various transitions in individual membrane potentials in $N$ $(=10^3)$ globally-coupled excitatory subthreshold Izhikevich neurons for $I_{DC}=3.6$ and $D=3.0$. Transition from spiking to bursting in individual potentials: time series of the membrane potential $v_1$ and the recovery variable $u_1$ of the 1st neuron for $ J=$ (a1) 0.6, (a2) 0.75, (a3) 0.8. Plots of fraction $n$ (number of spikes per burst) versus n for $J=$ (b1) 0.75 and (b2) 0.8. Average number of spikes $\overline n$ per burst versus $J$ is shown in (c). Transition from bursting to fast spiking in individual potentials: time series of the membrane potential $v_1$ for $J=$ (d1) 6.06, (d2) 6.07, and (d3) 6.09. Inverse of average burst length, $<BL>^{-1}$, versus $J$ is shown in (e). Transition from fast spiking to oscillator death in individual potentials: time series of the membrane potential $v_1$ for $J=$ (d1) 15, (d2) 18, and (d3) 20. Mean firing frequency $f$ versus $J$ is shown in (g).
}
\label{fig:OD}
\end{figure}

With increasing $J$, change in firing patterns of individual neurons and the corresponding population states are discussed. For small $J$ individual neurons exhibit noise-induced spikings. Figure \ref{fig:OD}(a1) shows the time series of the membrane potential $v_1$ and the recovery variable $u_1$ of the 1st neuron for $J=0.6$. Here, the slow variable $u_1$ provides a negative feedback to the fast variable $v_1$. Spiking $v_1$ pushes $u_1$ outside the spiking area. Then, $u_1$ slowly decays into the quiescent area [see Fig.~\ref{fig:OD}(a1)], which results in termination of spiking. This quiescent $v_1$ pushes $u_1$ outside the quiescent area; then, $u_1$ revisits the spiking area, which leads to spiking of $v_1$. Through repetition of this process, spikings appear successively in $v_1$, as shown in Fig.~\ref{fig:OD}(a1). Population synchronization between these individual spikings appear for $J_l^* < J < J_b^*$. However, as J passes a threshold $J_b^*$, the coherent synaptic input into the first neuron becomes so strong that a tendency that a spike in $v_1$ cannot push $u_1$ outside the spiking area occurs. As an example, see the case of J = 0.75 in Fig.~\ref{fig:OD}(a2). For this case, both spikings (singlets) and burstings (doublets consisting of two spikes) appear, as shown in Fig.~4(a2); 69 percentage of firings are singlets, while 31 percentages of firings are doublets [see Fig.~\ref{fig:OD}(b1)]. For $J=0.75$, after the 2nd spike in $v_1$, $u_1$ at first decreases a little (with nearly zero slope) and then increases abruptly up to a peak value of $u_1$, which is larger than that of $u_1$ for $J = 0.6$. Thus, after the 2nd spike, $u_1$ remains inside the spiking area; hence, a third spike, constituting a doublet, appears in $v_1$. After this 3rd spike, $u_1$ is pushed away from the spiking area and slowly decays into the quiescent area, which results in the termination of repetitive spikings. In this way, doublets appear in $v_1$ for J = 0.75. As J is further increased, the coherent synaptic input becomes stronger, so the number of spikes in a burst increases [e.g., see the doublets and triplets for $J = 0.8$ in Fig.~\ref{fig:OD}(a3)]; 88.7 percentage of firings are doublets, while 11.3 percentages of firings are triplets [see Fig.~\ref{fig:OD}(b2)]. Figure \ref{fig:OD}(c) shows the average number of spikes $\langle n \rangle$ per burst versus $J$, and $\langle n \rangle$ becomes larger than unity (i.e., burstings appear) for $J > J_b^*$. Population synchronization between these burstings occurs for $J_b^* < J < J_{fs}^*$. With increase in $J$, longer burst lengths (i.e., lengths of the active phase for the bursting activity) appear as shown in Figs.~\ref{fig:OD}(d1) and \ref{fig:OD}(d2), and eventually the  average burst length, $\langle BL \rangle$, diverges to the infinity (i.e., its inverse, ${\langle BL \rangle}^{-1}$, decreases to zero) as $J$ goes to $J_{fs}^*$ [see Fig.~\ref{fig:OD}(e)]. Then, for $J>J_{fs}^*$ individual neurons exhibit fast spikings as shown in Fig.~\ref{fig:OD}(d) for $J=6.09$. Since these fast spikes keep no pace with each other, incoherent states appear as shown in Fig.~\ref{fig:Order}(c). However, as $J$ is further increased, the coupling induces fast spike synchronization in a range of $J_{h1}^* < J < J_{h2}^*$. Then, slow spikings with longer spiking phases appear, as shown in Figs.~\ref{fig:OD}(f1) and \ref{fig:OD}(f2). Figure \ref{fig:OD}(g) shows the mean firing frequency $f$ (i.e., the inverse of the average interspike interval) versus the coupling strength. As $J$ approaches a threshold $J_{od}^*$, $f$ goes to zero. Consequently, for $J> J^*_{od}$ stochastic oscillator death (i.e., quenching of noise-induced slow spikings) occurs [e.g., see Fig.~\ref{fig:OD}(f3) for $J=20$].

\begin{figure}
\includegraphics[width=0.7\columnwidth]{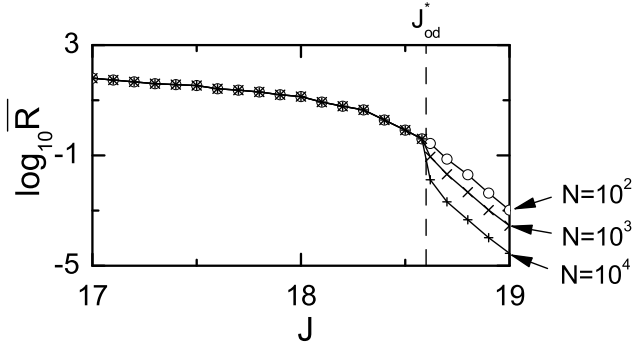}
\caption{Transition between firing and non-firing states in $N$ globally-coupled excitatory subthreshold Izhikevich neurons
for $I_{DC}=3.6$ and $D=3.0$. Plots of the average population spike rate $\overline{R}$ versus the coupling strength for $N=10^2$, $10^3$, and $10^4$.
}
\label{fig:Nonfiring}
\end{figure}

The stochastic oscillator death of individual neurons leads to a transition from firing to non-firing states at the population level. This firing-nonfiring transition may be well described in terms of the average population spike rate $\overline R$ which is a time average of the instantaneous population spike rate. To get a smooth instantaneous population spike rate $R(t)$, each spike in the raster plot is convoluted with a Gaussian kernel function \citep{KDE}:
\begin{equation}
 R(t) = {\frac {1} {N}} \sum_{i=1}^N \sum_{s=1}^{n_i} K_h(t-t_s^{(i)}),
\end{equation}
where $i$ is the neuron index, $t_s^{(i)}$ is the $s$th spike of the $i$th neuron, $n_i$ is the total number of spikes for the $i$th neuron, $N$ is the total number of neurons, and the Gaussian kernel of band width $h$ (=1 ms) is given by
\begin{equation}
K_h(t) = \frac {1} {\sqrt{2 \pi} h} e^{-t^2/2h^2}.
\end{equation}
Here, we discard the first time steps of a stochastic trajectory as transients during $10^3$ ms, and then we numerically compute $\overline R$ by following the stochastic trajectory for $3 \times10^4$ ms when $N=10^2, 10^3,$ and $10^4$. For the firing (non-firing) state, the average population spike rate $\overline R$ approaches a non-zero (zero) limit value in the thermodynamic limit of $N \rightarrow \infty$. Figure \ref{fig:Nonfiring} shows a plot of the average population spike rate versus the coupling strength. For $J > J_{od}^*$, $\overline R$ tends to zero as $N$ goes to the infinity, and hence non-firing states appear due to the stochastic oscillator death of individual neurons.

\section{Summary}
\label{sec:SUM}
We have studied coupling-induced population synchronization which may be used for efficient cognitive processing by changing the coupling strength $J$ in an excitatory population of subthreshold Izhikevich neurons. As $J$ is increased, rich population states have appeared in the following order: incoherent state $\rightarrow$ spike synchronization $\rightarrow$ burst synchronization $\rightarrow$ incoherent state $\rightarrow$ fast spike synchronization $\rightarrow$ incoherent state $\rightarrow$ non-firing state. Particularly, three types of population synchronization (i.e., spike, burst, and fast spike synchronization) have been found to occur. Transitions between population synchronization and incoherence have been well described in terms of a thermodynamic order parameter. These various transitions between population states have occurred due to emergence of the following diverse individual states: spiking $\rightarrow$ bursting $\rightarrow$ fast spiking $\rightarrow$ slow spiking $\rightarrow$ oscillator death. Each population synchronization and individual state were well characterized by using the techniques of nonlinear dynamics such as the raster plot of spikes, the time series of membrane potentials, and the phase portrait. As a final step, stochastic oscillator death (cessation of individual noise-induced slow spikings) occurred because each individual neuron is attracted to a noisy equilibrium state. This stochastic oscillator death leads to a transition from firing to non-firing states at the population level. The firing-nonfiring transition has also been characterized in terms of the average population spike rate. Since the Izhikevich model we employed for our study is a canonical model \citep{Burst1,Izhi1,Izhi2,Izhi3}, we expect that our results are still valid in other neuronal models. Finally, we note that population synchronization of noise-induced firings may lead to emergence of synchronous brain rhythms in a noisy environment which contribute to cognitive functions such as sensory perception, multisensory integration, selective attention, and working memory.

\end{document}